# THE RISE OF GRAPHENE

A.K. Geim and K.S. Novoselov
Manchester Centre for Mesoscience and Nanotechnology,
University of Manchester, Oxford Road M13 9PL, United Kingdom

**Graphene is a rapidly rising star on the horizon of materials science and condensed matter physics. This strictly two-dimensional material exhibits exceptionally high crystal and electronic quality and, despite its short history, has already revealed a cornucopia of new physics and potential applications, which are briefly discussed here. Whereas one can be certain of the realness of applications only when commercial products appear, graphene no longer requires any further proof of its importance in terms of fundamental physics. Owing to its unusual electronic spectrum, graphene has led to the emergence of a new paradigm of "relativistic" condensed matter physics, where quantum relativistic phenomena, some of which are unobservable in high energy physics, can now be mimicked and tested in table-top experiments. More generally, graphene represents a conceptually new class of materials that are only one atom thick and, on this basis, offers new inroads into low-dimensional physics that has never ceased to surprise and continues to provide a fertile ground for applications.**

Graphene is the name given to a flat monolayer of carbon atoms tightly packed into a two-dimensional (2D) honeycomb lattice, and is a basic building block for graphitic materials of all other dimensionalities (Figure 1). It can be wrapped up into 0D fullerenes, rolled into 1D nanotubes or stacked into 3D graphite. Theoretically, graphene (or "2D graphite") has been studied for sixty years[1-3] and widely used for describing properties of various carbon-based materials. Forty years later, it was realized that graphene also provides an excellent condensed-matter analogue of (2+1)-dimensional quantum electrodynamics[4-6], which propelled graphene into a thriving theoretical toy model. On the other hand, although known as integral part of 3D materials, graphene was presumed not to exist in the free state, being described as an "academic" material[5] and believed to be unstable with respect to the formation of curved structures such as soot, fullerenes and nanotubes. All of a sudden, the vintage model turned into reality, when free-standing graphene was unexpectedly found three years ago[7,8] and, especially, when the follow-up experiments[9,10] confirmed that its charge carriers were indeed massless Dirac fermions. So, the graphene "gold rush" has begun.

MATERIALS THAT SHOULD NOT EXIST
More than 70 years ago, Landau and Peierls argued that strictly two-dimensional (2D) crystals were thermodynamically unstable and could not exist[11,12]. Their theory pointed out that a divergent contribution of thermal fluctuations in low-dimensional crystal lattices should lead to such displacements of atoms that they become comparable to interatomic distances at any finite temperature[13]. The argument was later extended by Mermin[14] and is strongly supported by a whole omnibus of experimental observations. Indeed, the melting temperature of thin films rapidly decreases with decreasing thickness, and they become unstable (segregate into islands or decompose) at a thickness of, typically, dozens of atomic layers[15,16]. For this reason, atomic monolayers have so far been known only as an integral part of larger 3D structures, usually grown epitaxially on top of monocrystals with matching crystal lattices[15,16]. Without such a 3D base, 2D materials were presumed not to exist until 2004, when the common wisdom was flaunted by the experimental discovery of graphene[7] and other free-standing 2D atomic crystals (for example, single-layer boron nitride and half-layer BSCCO)[8]. These crystals could be obtained on top of non-crystalline substrates[8-10], in liquid suspension[7,17] and as suspended membranes[18].

Importantly, the 2D crystals were found not only to be continuous but to exhibit high crystal quality[7-10,17,18]. The latter is most obvious for the case of graphene, in which charge carriers can travel thousands interatomic distances without scattering[7-10]. With the benefit of hindsight, the existence of such one-atom-thick crystals can be reconciled with theory. Indeed, it can be argued that the obtained 2D crystallites are quenched in a metastable state because they are extracted from 3D materials, whereas their small size (<<1mm) and strong interatomic bonds assure that thermal fluctuations cannot lead to the generation of dislocations or other crystal defects even at elevated temperature[13,14].

A complementary viewpoint is that the extracted 2D crystals become intrinsically stable by gentle crumpling in the third dimension on a lateral scale of ≈10nm[18,19]. Such 3D warping observed experimentally[18] leads to a gain in elastic energy but suppresses thermal vibrations (anomalously large in 2D), which above a certain temperature can minimize the total free energy[19].

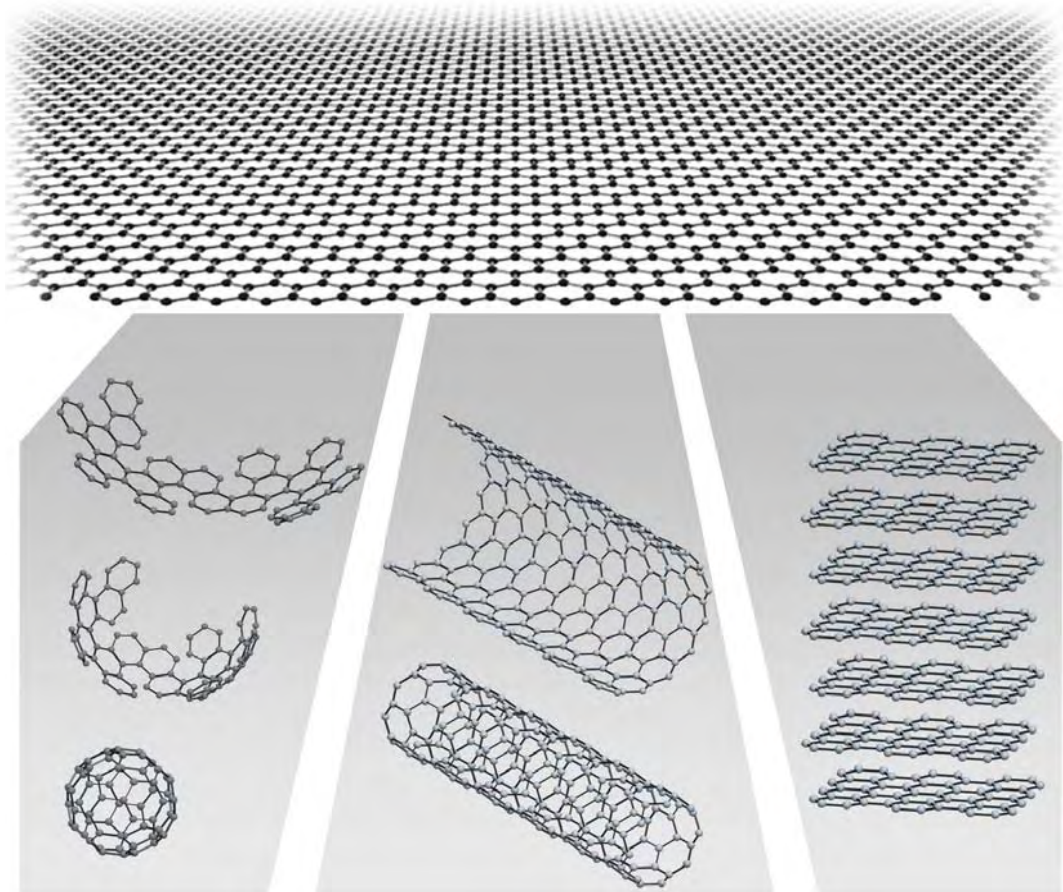

Figure 1. **Mother of all graphitic forms.** Graphene is a 2D building material for carbon materials of all other dimensionalities. It can be wrapped up into 0D buckyballs, rolled into 1D nanotubes or stacked into 3D graphite.

BRIEF HISTORY OF GRAPHENE
Before reviewing the earlier work on graphene, it is useful to define what 2D crystals are. Obviously, a single atomic plane *is* a 2D crystal, whereas 100 layers should be considered as a thin film of a 3D material. But how many layers are needed to make a 3D structure? For the case of graphene, the situation has recently become reasonably clear. It was shown that the electronic structure rapidly evolves with the number of layers, approaching the 3D limit of graphite already at 10 layers[20]. Moreover, only graphene and, to a good approximation, its bilayer have simple electronic spectra: they are both zero-gap semiconductors (can also be referred to as zero-overlap semimetals) with one type of electrons and one type of holes. For 3 and more layers, the spectra become increasingly complicated: Several charge carriers appear[7,21], and the conduction and valence bands start notably overlapping[7,20]. This allows one to distinguish between single-, double- and few- (3 to <10) layer graphene as three different types of 2D crystals ("graphenes"). Thicker structures should be considered, to all intents and purposes, as thin films of graphite. From the experimental point of view, such a definition is also sensible. The screening length in graphite is only ≈5Å (that is, less than 2 layers in thickness)[21] and, hence, one must differentiate between the surface and the bulk even for films as thin as 5 layers.[21,22]

Earlier attempts to isolate graphene concentrated on chemical exfoliation. To this end, bulk graphite was first intercalated (to stage I)[23] so that graphene planes became separated by layers of intervening atoms or molecules.

This usually resulted in new 3D materials[23]. However, in certain cases, large molecules could be inserted between atomic planes, providing greater separation such that the resulting compounds could be considered as isolated graphene layers embedded in a 3D matrix. Furthermore, one can often get rid of intercalating molecules in a chemical reaction to obtain a sludge consisting of restacked and scrolled graphene sheets[24-26]. Because of its uncontrollable character, graphitic sludge has so far attracted only limited interest.

There have also been a small number of attempts to grow graphene. The same approach as generally used for growth of carbon nanotubes so far allowed graphite films only thicker than $\approx 100$ layers[27]. On the other hand, single- and few-layer graphene have been grown epitaxially by chemical vapour deposition of hydrocarbons on metal substrates[28,29] and by thermal decomposition of SiC[30-34]. Such films were studied by surface science techniques, and their quality and continuity remained unknown. Only lately, few-layer graphene obtained on SiC was characterized with respect to its electronic properties, revealing high-mobility charge carriers[32,33]. Epitaxial growth of graphene offers probably the only viable route towards electronic applications and, with so much at stake, a rapid progress in this direction is expected. The approach that seems promising but has not been attempted yet is the use of the previously demonstrated epitaxy on catalytic surfaces[28,29] (such as Ni or Pt) followed by the deposition of an insulating support on top of graphene and chemical removal of the primary metallic substrate.

THE ART OF GRAPHITE DRAWING
In the absence of quality graphene wafers, most experimental groups are currently using samples obtained by micromechanical cleavage of bulk graphite, the same technique that allowed the isolation of graphene for the first time[7,8]. After fine-tuning, the technique[8] now provides high-quality graphene crystallites up to 100 μm in size, which is sufficient for most research purposes (see Figure 2). Superficially, the technique looks as nothing more sophisticated than drawing by a piece of graphite[8] or its repeated peeling with adhesive tape[7] until the thinnest flakes are found. A similar approach was tried by other groups (earlier[35] and independently[22,36]) but only graphite flakes 20 to 100 layers thick were found. The problem is that graphene crystallites left on a substrate are extremely rare and hidden in a "haystack" of thousands thick (graphite) flakes. So, even if one were deliberately searching for graphene by using modern techniques for studying atomically thin materials, it would be impossible to find those several micron-size crystallites dispersed over, typically, a 1-cm$^2$ area. For example, scanning-probe microscopy has too low throughput to search for graphene, whereas scanning electron microscopy is unsuitable because of the absence of clear signatures for the number of atomic layers.

The critical ingredient for success was the observation[7,8] that graphene becomes visible in an optical microscope if placed on top of a Si wafer with a carefully chosen thickness of $SiO_2$, owing to a feeble interference-like contrast with respect to an empty wafer. If not for this simple yet effective way to scan substrates in search of graphene crystallites, they would probably remain undiscovered today. Indeed, even knowing the exact recipe[7,8], it requires special care and perseverance to find graphene. For example, only a 5% difference in $SiO_2$ thickness (315 nm instead of the current standard of 300 nm) can make single-layer graphene completely invisible. Careful selection of the initial graphite material (so that it has largest possible grains) and the use of freshly -cleaved and -cleaned surfaces of graphite and $SiO_2$ can also make all the difference. Note that graphene was recently[37,38] found to have a clear signature in Raman microscopy, which makes this technique useful for quick thickness inspection, even though potential crystallites still have to be first hunted for in an optical microscope.

Similar stories could be told about other 2D crystals (particularly, dichalcogenides monolayers) where many attempts were made to split these strongly layered materials into individual planes[39,40]. However, the crucial step of isolating monolayers to assess their properties individually was never achieved. Now, by using the same approach as demonstrated for graphene, it is possible to investigate potentially hundreds of different 2D crystals[8] in search of new phenomena and applications.

FERMIONS GO BALLISTIC
Although there is a whole class of new 2D materials, all experimental and theoretical efforts have so far focused on graphene, somehow ignoring the existence of other 2D crystals. It remains to be seen whether this bias is justified but the primary reason for it is clear: It is the exceptional electronic quality exhibited by the isolated graphene crystallites[7-10]. From experience, people know that high-quality samples always yield new physics, and this understanding has played a major role in focusing attention on graphene.

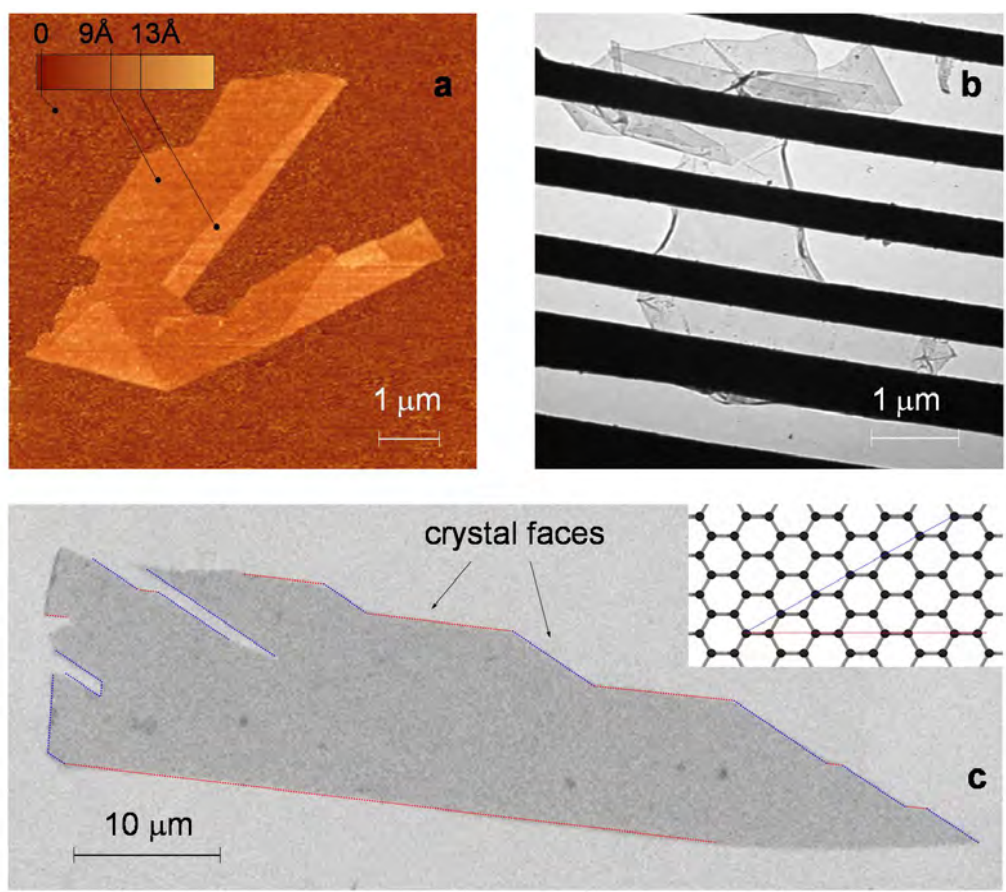

Figure 2. **One-atom-thick single crystals: the thinnest material you will ever see**. **a**, Graphene visualized by atomic-force microscopy (adapted from ref. 8). The folded region exhibiting a relative height of ≈4Å clearly indicates that it is a single layer. **b**, A graphene sheet freely suspended on a micron-size metallic scaffold. The transmission-electron-microscopy image is adapted from ref. 18. **c**, scanning-electron micrograph of a relatively large graphene crystal, which shows that most of the crystal's faces are zigzag and armchair edges as indicated by blue and red lines and illustrated in the inset (*T.J. Booth, K.S.N, P. Blake & A.K.G.* unpublished). 1D transport along zigzag edges and edge-related magnetism are expected to attract significant attention.

Graphene's quality clearly reveals itself in a pronounced ambipolar electric field effect (Fig. 3a) such that charge carriers can be tuned continuously between electrons and holes in concentrations $n$ as high as $10^{13} cm^{-2}$ and their mobilities $\mu$ can exceed 15,000 $cm^2/Vs$ even under ambient conditions[7-10]. Moreover, the observed mobilities weakly depend on temperature $T$, which means that $\mu$ at 300K is still limited by impurity scattering and, therefore, can be improved significantly, perhaps, even up to ≈100,000 $cm^2/Vs$. Although some semiconductors exhibit room-temperature $\mu$ as high as ≈77,000 $cm^2/Vs$ (namely, InSb), those values are quoted for undoped bulk semiconductors. In graphene, $\mu$ remains high even at high $n$ (>$10^{12} cm^{-2}$) in both electrically- and chemically- doped devices[41], which translates into ballistic transport on submicron scale (up to ≈0.3 μm at 300K). A further indication of the system's extreme electronic quality is the quantum Hall effect (QHE) that can be observed in graphene even at room temperature (Fig. 3b), extending the previous temperature range for the QHE by a factor of 10.

An equally important reason for the interest in graphene is a unique nature of its charge carriers. In condensed matter physics, the Schrödinger equation rules the world, usually being quite sufficient to describe electronic properties of materials. Graphene is an exception: Its charge carriers mimic relativistic particles and are easier and more natural to describe starting with the Dirac equation rather than the Schrödinger equation[4-6,42-47]. Although there is nothing particularly relativistic about electrons moving around carbon atoms, their interaction with a periodic potential of graphene's honeycomb lattice gives rise to new quasiparticles that at low energies $E$ are accurately described by the (2+1)-dimensional Dirac equation with an effective speed of light $v_F \approx 10^6$ m/s. These quasiparticles, called massless Dirac fermions, can be seen as electrons that lost their rest mass $m_0$ or as neutrinos that acquired the electron charge $e$. The relativistic-like description of electron waves on honeycomb lattices has been known theoretically for many years, never failing to attract attention, and the experimental discovery of graphene now provides a way to probe quantum electrodynamics (QED) phenomena by measuring graphene's electronic properties.

QED IN A PENCIL TRACE

From the point of view of its electronic properties, graphene is a zero-gap semiconductor, in which low-$E$ quasiparticles within each valley can formally be described by the Dirac-like Hamiltonian

$$\hat{H} = \hbar v_F \begin{pmatrix} 0 & k_x - ik_y \\ k_x + ik_y & 0 \end{pmatrix} = \hbar v_F \vec{\sigma} \cdot \vec{k}$$

where $\vec{k}$ is the quasiparticle momentum, $\vec{\sigma}$ the 2D Pauli matrix and the $k$-independent Fermi velocity $v_F$ plays the role of the speed of light. The Dirac equation is a direct consequence of graphene's crystal symmetry. Its honeycomb lattice is made up of two equivalent carbon sublattices $A$ and $B$, and cosine-like energy bands associated with the sublattices intersect at zero $E$ near the edges of the Brillouin zone, giving rise to conical sections of the energy spectrum for $|E| <$ 1eV (Fig. 3).

We emphasize that the linear spectrum $E = \hbar v_F k$ is not the only essential feature of the band structure. Indeed, electronic states near zero $E$ (where the bands intersect) are composed of states belonging to the different sublattices, and their relative contributions in quasiparticles' make-up have to be taken into account by, for example, using two-component wavefunctions (spinors). This requires an index to indicate sublattices $A$ and $B$, which is similar to the spin index (up and down) in QED and, therefore, is referred to as pseudospin. Accordingly, in the formal description of graphene's quasiparticles by the Dirac-like Hamiltonian above, $\vec{\sigma}$ refers to pseudospin rather than the real spin of electrons (the latter must be described by additional terms in the Hamiltonian). Importantly, QED-specific phenomena are often inversely proportional to the speed of light $c$ and, therefore, enhanced in graphene by a factor $c/v_F \approx 300$. In particular, this means that pseudospin-related effects should generally dominate those due to the real spin.

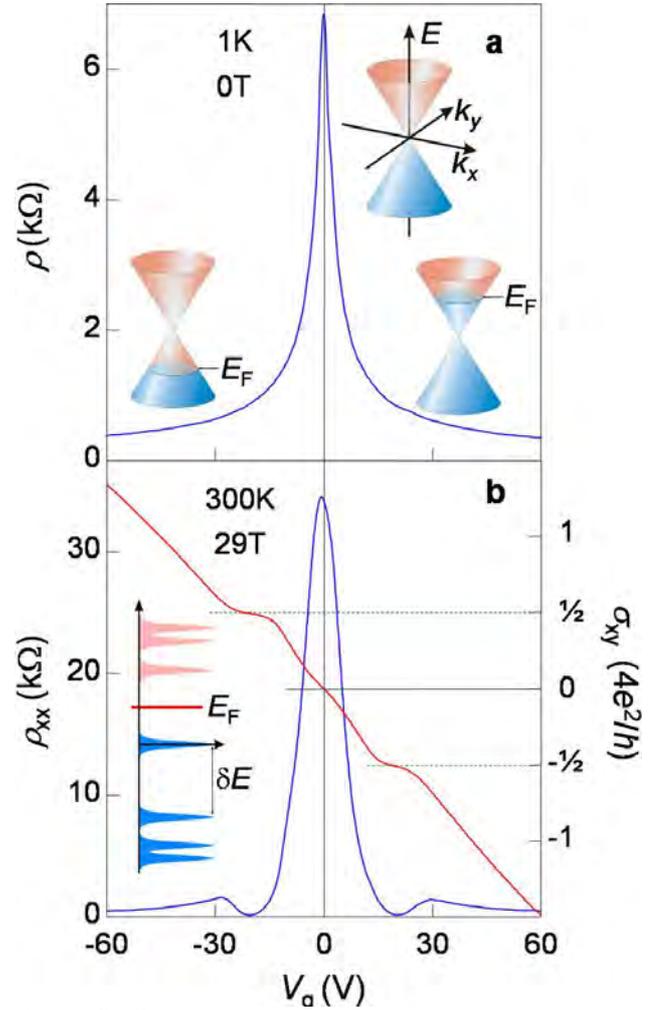

Figure 3. **Ballistic electron transport in graphene. a**, Ambipolar electric field effect in single-layer graphene. The insets show its conical low-energy spectrum $E(k)$, indicating changes in the position of the Fermi energy $E_F$ with changing gate voltage $V_g$. Positive (negative) $V_g$ induce electrons (holes) in concentrations $n = \alpha V_g$ where the coefficient $\alpha \approx 7.2 \cdot 10^{10}$ cm$^{-2}$/V for field-effect devices with a 300 nm SiO$_2$ layer used as a dielectric[7-9]. The rapid decrease in resistivity $\rho$ with adding charge carriers indicates their high mobility (in this case, $\mu \approx 5{,}000$cm$^2$/Vs and does not noticeably change with temperature up to 300K). **b**, Room-temperature quantum Hall effect (*K.S.N., Z. Jiang, Y. Zhang, S.V. Morozov, H.L. Stormer, U. Zeitler, J.C. Maan, G.S. Boebinger, P. Kim & A.K.G. Science* 2007, in the press). Because quasiparticles in graphene are massless and also exhibit little scattering even under ambient conditions, the QHE survives up to room $T$. Shown in red is the Hall conductivity $\sigma_{xy}$ that exhibits clear plateaux at $2e^2/h$ for both electrons and holes. The longitudinal conductivity $\rho_{xx}$ (blue) reaches zero at the same gate voltages. The inset illustrates the quantized spectrum of graphene where the largest cyclotron gap is described by $\delta E$(K)$\approx 420 \cdot \sqrt{B}$ (T).

By analogy with QED, one can also introduce a quantity called chirality[6] that is formally a projection of $\vec{\sigma}$ on the direction of motion $\vec{k}$ and is positive (negative) for electrons (holes). In essence, chirality in graphene signifies the fact that $k$ electron and $-k$ hole states are intricately connected by originating from the same carbon sublattices. The concepts of chirality and pseudospin are important because many electronic processes in graphene can be understood as due to conservation of these quantities.[6,42-47]

It is interesting to note that in some narrow-gap 3D semiconductors, the gap can be closed by compositional changes or by applying high pressure. Generally, zero gap does not necessitate Dirac fermions (that imply conjugated electron and hole states) but, in some cases, they may appear[5]. The difficulties of tuning the gap to zero, while keeping carrier mobilities high, the lack of possibility to control electronic properties of 3D materials by the electric field effect and, generally, less pronounced quantum effects in 3D limited studies of such semiconductors mostly to measuring the concentration dependence of their effective masses $m$ (for a review, see ref. 48). It is tempting to have a fresh look at zero-gap bulk semiconductors, especially because Dirac fermions were recently reported even in such a well-studied (small-overlap) 3D material as graphite.[49,50]

CHIRAL QUANTUM HALL EFFECTS
At this early stage, the main experimental efforts have been focused on electronic properties of graphene, trying to understand the consequences of its QED-like spectrum. Among the most spectacular phenomena reported so far, there are two new ("chiral") quantum Hall effects, minimum quantum conductivity in the limit of vanishing concentrations of charge carriers and strong suppression of quantum interference effects.

Figure 4 shows three types of the QHE behaviour observed in graphene. The first one is a relativistic analogue of the integer QHE and characteristic to single-layer graphene[9,10]. It shows up as an uninterrupted ladder of equidistant steps in Hall conductivity $\sigma_{xy}$ which persists through the neutrality (Dirac) point, where charge carriers change from electrons to holes (Fig. 4a). The sequence is shifted with respect to the standard QHE sequence by ½, so that $\sigma_{xy} = \pm 4e^2/h\,(N + \tfrac{1}{2})$ where $N$ is the Landau level (LL) index and factor 4 appears due to double valley and double spin degeneracy. This QHE has been dubbed "half-integer" to reflect both the shift and the fact that, although it is not a new fractional QHE, it is not the standard integer QHE either. The unusual sequence is now well understood as arising due to the QED-like quantization of graphene's electronic spectrum in magnetic field *B*, which is described[44,51-53] by $E_N = \pm v_F \sqrt{2e\hbar BN}$ where sign ± refers to electrons and holes. The existence of a quantized level at zero *E*, which is shared by electrons and holes (Fig. 4c), is essentially everything one needs to know to explain the anomalous QHE sequence.[51-55] An alternative explanation for the half-integer QHE is to invoke the coupling between pseudospin and orbital motion, which gives rise to a geometrical phase of π accumulated along cyclotron trajectories and often referred to as Berry's phase.[9,10,56] The additional phase leads to a π-shift in the phase of quantum oscillations and, in the QHE limit, to a half-step shift.

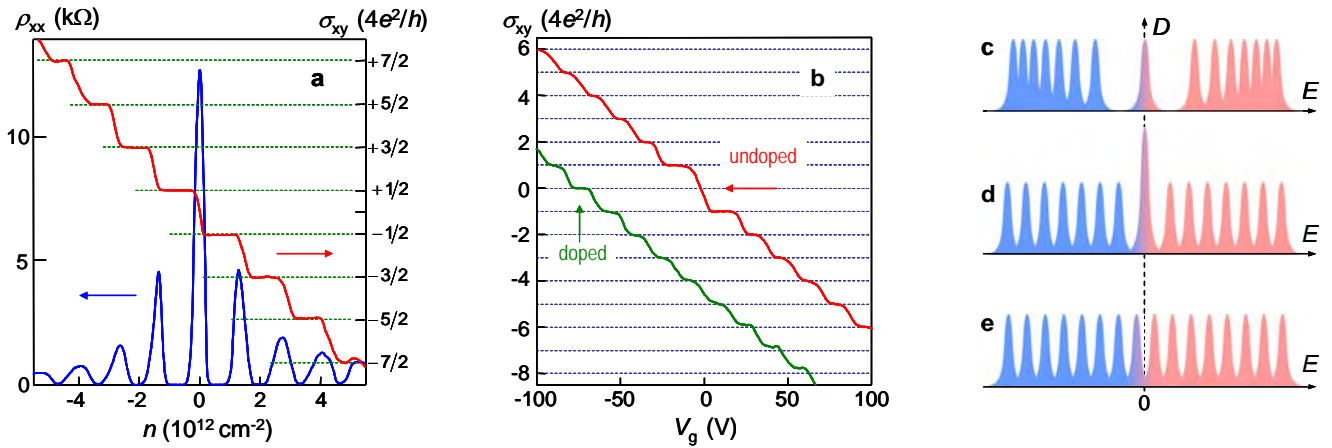

Figure 4. **Chiral quantum Hall effects.** **a**, The hallmark of massless Dirac fermions is QHE plateaux in $\sigma_{xy}$ at half integers of $4e^2/h$ (adapted from ref. 9). **b**, Anomalous QHE for massive Dirac fermions in bilayer graphene is more subtle (red curve[55]): $\sigma_{xy}$ exhibits the standard QHE sequence with plateaux at all integer $N$ of $4e^2/h$ except for $N=0$. The missing plateau is indicated by the red arrow. The zero-$N$ plateau can be recovered after chemical doping, which shifts the neutrality point to high $V_g$ so that an asymmetry gap (≈0.1eV in this case) is opened by the electric field effect (green curve; adapted from ref. 59). **c-e**, Different types of Landau quantization in graphene. The sequence of Landau levels in the density of states $D$ is described by $E_N \propto \sqrt{N}$ for *massless* Dirac fermions in single-layer graphene (**c**) and by $E_N \propto \sqrt{N(N-1)}$ for *massive* Dirac fermions in bilayer graphene (**d**). The standard LL sequence $E_N \propto (N+\tfrac{1}{2})$ is expected to recover if an electronic gap is opened in the bilayer (**e**).

Bilayer graphene exhibits an equally anomalous QHE (Fig 4b)[55]. Experimentally, it shows up less spectacular: One measures the standard sequence of Hall plateaux $\sigma_{xy} = \pm N\, 4e^2/h$ but the very first plateau at $N=0$ is missing, which also implies that bilayer graphene remains metallic at the neutrality point.[55] The origin of this anomaly lies in a rather bizarre nature of quasiparticles in bilayer graphene, which are described[57] by

$$\hat{H} = -\frac{\hbar^2}{2m}\begin{pmatrix} 0 & (k_x - ik_y)^2 \\ (k_x + ik_y)^2 & 0 \end{pmatrix}$$

This Hamiltonian combines the off-diagonal structure, similar to the Dirac equation, with Schrödinger-like terms $\hat{p}^2/2m$. The resulting quasiparticles are chiral, similar to massless Dirac fermions, but have a finite mass *m* ≈0.05$m_0$. Such massive chiral particles would be an oxymoron in relativistic quantum theory. The Landau quantization of "massive Dirac fermions" is given[57] by $E_N = \pm \hbar \omega_c \sqrt{N(N-1)}$ with two degenerate levels $N=0$ and 1 at zero *E* ($\omega_c$ is the cyclotron frequency). This additional degeneracy leads to the missing zero-*E* plateau and the double-height step in Fig. 4b. There is also a pseudospin associated with massive Dirac fermions, and its orbital rotation leads to a geometrical phase of 2π. This phase is indistinguishable from zero in the quasiclassical limit ($N \gg 1$) but reveals itself in the double degeneracy of the zero-*E* LL (Fig. 4d).[55]

It is interesting that the "standard" QHE with all the plateaux present can be recovered in bilayer graphene by the electric field effect (Fig. 4b). Indeed, gate voltage not only changes *n* but simultaneously induces an asymmetry between the two graphene layers, which results in a semiconducting gap[58,59]. The electric-field-induced gap eliminates the additional degeneracy of the zero-*E* LL and leads to the uninterrupted QHE sequence by splitting the double step into two (Fig. 4e)[58,59]. However, to observe this splitting in the QHE measurements, one needs to probe the region near the neutrality point at finite $V_g$, which can be achieved by additional chemical doping[59]. Note that bilayer graphene is the only known material in which the electronic band structure changes significantly by the electric field effect and the semiconducting gap *ΔE* can be tuned continuously from zero to ≈0.3eV if SiO$_2$ is used as a dielectric.

CONDUCTIVITY "WITHOUT" CHARGE CARRIERS

Another important observation is that graphene's zero-field conductivity σ does not disappear in the limit of vanishing *n* but instead exhibits values close to the conductivity quantum $e^2/h$ per carrier type[9]. Figure 5 shows the lowest conductivity $\sigma_{min}$ measured near the neutrality point for nearly 50 single-layer devices. For all other known materials, such a low conductivity unavoidably leads to a metal-insulator transition at low *T* but no sign of the transition has been observed in graphene down to liquid-helium *T*. Moreover, although it is the persistence of the metallic state with σ of the order of $e^2/h$ that is most exceptional and counterintuitive, a relatively small spread of the observed conductivity values (see Fig. 5) also allows one to speculate about the quantization of $\sigma_{min}$. We emphasize that it is the resistivity (conductivity) that is quantized in graphene, in contrast to the resistance (conductance) quantization known in many other transport phenomena.

Minimum quantum conductivity has been predicted for Dirac fermions by a number of theories[5,44,45,47,60-64]. Some of them rely on a vanishing density of states at zero *E* for the linear 2D spectrum. However, comparison between the experimental behaviour of massless and massive Dirac fermions in graphene and its bilayer allows one to distinguish between chirality- and masslessness- related effects. To this end, bilayer graphene also exhibits a minimum conductivity of the order of $e^2/h$ per carrier type,[55,65] which indicates that it is chirality, rather than the linear spectrum, that is more important. Most theories suggest $\sigma_{min} = 4e^2/h\pi$, which is of about π times smaller than the typical values observed experimentally. One can see in Fig. 5 that the experimental data do not approach this theoretical value and mostly cluster around $\sigma_{min} = 4e^2/h$ (except for one low-μ sample that is rather unusual by also exhibiting 100%-normal weak localization behaviour at high *n*; see below). This disagreement has become known as "the mystery of a missing pie", and it remains unclear whether it is due to theoretical approximations about electron scattering in graphene or because the experiments probed only a limited range of possible sample parameters (e.g., length-to-width ratios[47]). To this end, note that close to the neutrality point ($n \leq 10^{11}$cm$^{-2}$) graphene is likely to conduct as a random network of electron and hole puddles (*A.K.G. & K.S.N.* unpublished). Such *microscopic* inhomogeneity is probably inherent to graphene (because of graphene sheet's warping/rippling)[18,66] but so far has not been taken into account by theory. Furthermore, *macroscopic* inhomogeneity (on the scale larger than the mean free path *l*) also plays an important role in measurements of $\sigma_{min}$. The latter inhomogeneity can explain a high-$\sigma$ tail in the data scatter in Fig. 5 by the fact that $\sigma$ reached its lowest values at slightly different $V_g$ in different parts of a sample, which yields effectively higher values of experimentally measured $\sigma_{min}$.

WEAK LOCALIZATION IN SHORT SUPPLY

At low temperatures, all metallic systems with high resistivity should inevitably exhibit large quantum-interference (localization) magnetoresistance, eventually leading to the metal-insulator transition at $\sigma \approx e^2/h$. Until now, such behaviour has been absolutely universal but it was found missing in graphene. Even near the neutrality point, no significant low-field ($B<1$T) magnetoresistance has been observed down to liquid-helium temperatures[66] and, although sub-100 nm Hall crosses did exhibit giant resistance fluctuations (S.V. Morozov, K.S.N., A.K.G. et al, unpublished), those could be attributed to changes in the distribution of electron and hole puddles and size quantization. It remains to be seen whether localization effects at the Dirac point recover at lower *T*, as the phase-breaking length becomes increasingly longer,[67] or the observed behaviour indicates a "marginal Fermi liquid"[68,43], in which the phase-breaking length goes to zero with decreasing *E*. Further experimental studies are much needed in this regime but it is difficult to probe because of microscopic inhomogeneity.

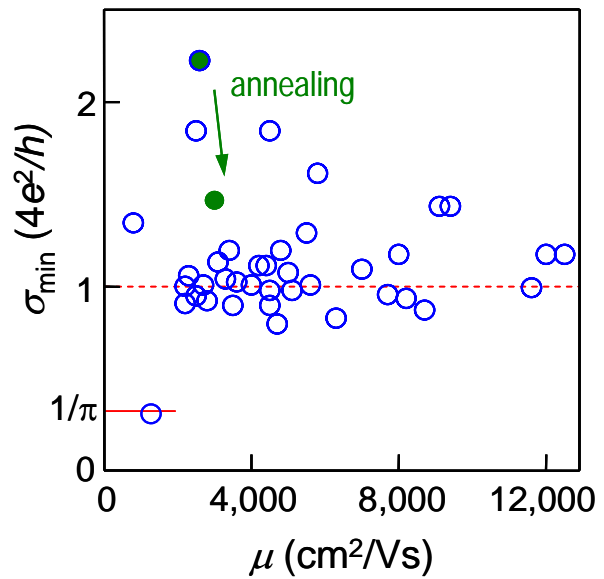

Figure 5. **Minimum conductivity of graphene.** Independent of their carrier mobility $\mu$, different graphene devices exhibited approximately the same conductivity at the neutrality point (open circles) with most data clustering around $\approx 4e^2/h$ indicated for clarity by the dashed line (*A.K.G. & K.S.N.* unpublished; includes the published data points from ref. 9). The high-conductivity tail is attributed to macroscopic inhomogeneity: by improving samples' homogeneity, $\sigma_{min}$ generally decreases, moving closer to $\approx 4e^2/h$. The green arrow and symbols show one of the devices that initially exhibited an anomalously large value of $\sigma_{min}$ but after thermal annealing at 400K its $\sigma_{min}$ moved closer to the rest of the statistical ensemble. Most of the data are taken in the bend resistance geometry where the macroscopic inhomogeneity plays the least role.

Away from the Dirac point (where graphene becomes a good metal), the situation has recently become reasonably clear. Universal conductance fluctuations (UCF) were reported to be qualitatively normal in this regime, whereas weak localization (WL) magnetoresistance was found to be somewhat random, varying for different samples from being virtually absent to showing the standard behaviour[66]. On the other hand, early theories had also predicted every possible type of WL magnetoresistance in graphene, from positive to negative to zero. Now it is understood that, for large *n* and in the absence of inter-valley scattering, there should be no magnetoresistance, because the triangular warping of graphene's Fermi surface destroys time-reversal symmetry *within* each valley.[69] With increasing inter-valley scattering, the normal (negative) WL should recover. Changes in inter-valley scattering rates by, for example, varying microfabrication procedures can explain the observed sample-dependent behaviour. A complementary explanation is that a sufficient inter-valley scattering is already present in the studied samples but the time-reversal symmetry is destroyed by elastic strain due to microscopic warping[66,70]. The strain in graphene has turned out to be equivalent to a random magnetic field, which also destroys time-reversal symmetry and suppresses WL. Whatever the mechanism, theory expects (approximately[71]) normal UCF at high *n*, in agreement with the experiment[66].

PENCILLED-IN BIG PHYSICS

Due to space limitations, we do not attempt to overview a wide range of other interesting phenomena predicted for graphene theoretically but as yet not observed experimentally. Nevertheless, let us mention two focal points for current theories. One of them is many-body physics near the Dirac point, where interaction effects should be strongly enhanced due to weak screening, the vanishing density of states and graphene's large coupling constant $e^2/\hbar v_F \approx 1$ ("effective fine structure constant"[68,72]). The predictions include various options for the fractional QHE, quantum Hall ferromagnetism, excitonic gaps, etc. (e.g., see refs. 44,72-79). The first relevant experiment in ultra-high $B$ has reported the lifting of spin and valley degeneracy[80].

Second, graphene is discussed in the context of testing various QED effects, among which the *gedanken* Klein paradox and *zitterbewegung* stand out because these effects are unobservable in particle physics. The notion of Klein paradox refers to a counterintuitive process of perfect tunnelling of relativistic electrons through arbitrarily high and wide barriers. The experiment is conceptually easy to implement in graphene[46]. *Zitterbewegung* is a term describing jittery movements of an electron due to interference between parts of its wavepacket belonging to positive (electron) and negative (positron) energy states. These quasi-random movements can be responsible for the finite conductivity $\approx e^2/h$ of ballistic devices,[45,47] are hypothesized to result in excess shot noise[47] and, maybe, can even be visualized by direct imaging[81,82] of Dirac trajectories. In the latter respect, graphene offers truly unique opportunities because, unlike in most semiconductor systems, its 2D electronic states are not buried deep under the surface and can be accessed directly by tunnelling and other local probes. One can expect many interesting results arising from scanning-tunnelling experiments in graphene. Another tantalizing possibility is to study QED in a curved space (by controllable bending of a graphene sheet) which allows one to address certain cosmological problems.[83]

2D OR NOT 2D

In addition to QED physics, there are many other reasons that should perpetuate active interest in graphene. For the sake of brevity, one can summarize them by referring to analogies with carbon nanotubes and 2D electron gases in semiconductors. Indeed, much of nanotubes' fame and glory can probably be credited to graphene, the very material they are made of. By projecting the accumulated knowledge about carbon nanotubes onto their flat counterpart and bearing in mind the rich physics brought about by semiconductor 2D systems, one can probably draw a reasonably good sketch of emerging opportunities.

The relationship between 2D graphene and 1D carbon nanotubes requires a special mentioning. The current rapid progress on graphene has certainly benefited from the relatively mature research on nanotubes that continue to provide a near-term guide in searching for graphene applications. However, there exists a popular opinion that graphene should be considered simply as unfolded carbon nanotubes and, therefore, can compete with them in the myriad of applications already suggested. Partisans of this view often claim that graphene will make nanotubes obsolete, allowing all the promised applications to reach an industrial stage because, unlike nanotubes, graphene can (probably) be produced in large quantities with fully reproducible properties. This view is both unfair and inaccurate. Dimensionality is one of the most defining material parameters and, as carbon nanotubes exhibit properties drastically different from those of 3D graphite and 0D fullerenes, 2D graphene is also quite different from its forms in the other dimensions. Depending on a particular problem in hand, graphene's prospects can be sometimes superior, sometimes inferior and, most often, completely different from those of carbon nanotubes or, for the sake of argument, of graphite.

"GRAPHENIUM INSIDE"

As concerns applications, graphene-based electronics should be mentioned first. This is because most efforts have so far been focused on this direction and such companies as Intel and IBM fund this research in order to keep an eye on possible developments. It is not surprising because, at the time when the Si-based technology is approaching its fundamental limits, any new candidate material to take over from Si is welcome, and graphene seems to offer an exceptional choice.

Graphene's potential for electronics is usually justified by citing high mobility of its charge carriers. However, as mentioned above, the truly exceptional feature of graphene is that its $\mu$ remains high even at highest electric-field-induced concentrations and seems to be little affected by chemical doping[41]. This translates into ballistic transport on a submicron scale at 300K. A room-temperature ballistic transistor has been a holy grail for electronic engineers, and graphene can make it happen. The large value of $v_F$ and low-resistance contacts without a Schottky barrier[7] should help further reducing the switching time perhaps to less than $10^{-13}$s. Relatively low on-off ratios (reaching only ≈100 because of graphene's minimum conductivity) do not seem to present a fundamental problem for high-frequency applications[7], and the demonstration of transistors operational at THz frequencies would be an important milestone for graphene-based electronics.

For mainstream logic applications, the fact that graphene remains metallic even at the neutrality point is a major problem. However, significant semiconductor gaps $\Delta E$ can still be engineered in graphene. As mentioned above, $\Delta E$ up to 0.3eV can be induced in bilayer graphene but this is perhaps more interesting in terms of tuneable infrared lasers and detectors. For single-layer graphene, $\Delta E$ can be induced by spatial confinement or lateral-superlattice potential. The latter seems to be a relatively straightforward solution because sizeable gaps (>0.1eV) should naturally occur in graphene epitaxially grown on top of crystals with a matching lattice such as boron nitride or the same SiC[30-34], in which large superlattice effects are undoubtedly expected.

Owing to graphene's linear spectrum and large $v_F$, the confinement gap is also rather large $\Delta E(\text{eV}) \approx \alpha \pi \hbar v_F / d \approx 1/d(\text{nm})$[84-86], as compared to other semiconductors, and it requires ribbons with width $d$ of about 10 nm for room-$T$ operation (coefficient $\alpha$ is ≈½ for Dirac fermions)[86]. With the Si-based technology rapidly advancing into this scale, the required size is no longer seen as a significant hurdle, and much research is expected along this direction. However, unless a technique for anisotropic etching of graphene is found to make devices with crystallographically defined faces (e.g., zigzag or armchair), one has to deal with conductive channels having irregular edges. In short channels, electronic states associated with such edges can induce a significant sample-dependent conductance.[84-86] In long channels, random edges may lead to additional scattering, which can be detrimental for transistors' speed and energy consumption and, in effect, cancel all the advantages offered by graphene's ballistic transport. Fortunately, high-anisotropy dry etching is probably achievable in graphene, owing to quite different chemical reactivity of zigzag and armchair edges.

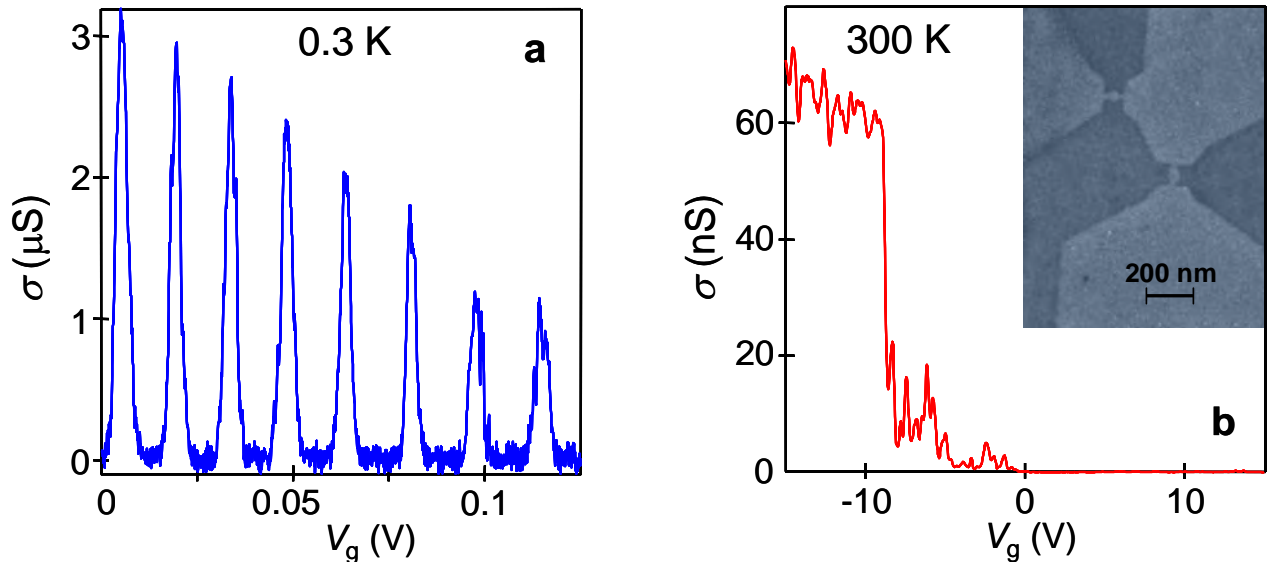

Figure 6. **Towards graphene-based electronics**. To achieve transistor action, nanometre ribbons and quantum dots can be carved in graphene (*L.A. Ponomarenko, F. Schedin, K.S.N. & A.K.G.*, in preparation). **a**, Coulomb blockade in relatively large quantum dots (diameter ≈0.25 μm) at low $T$. Conductance $\sigma$ of such devices can be controlled by either the back gate or a side electrode also made from graphene. Narrow constrictions in graphene with low-$T$ resistance much larger than 100kΩ serve as quantum barriers. **b**, 10-nm-scale graphene structures remain remarkably stable under ambient conditions and survive thermal cycling to liquid-helium $T$. Such devices show a high-quality transistor action even at room $T$ so that their conductance can be pinched-off completely over a large range of gate voltages near the neutrality point. The inset shows a scanning-electron micrograph of two graphene dots of ≈40 nm in diameter with narrower (<10nm) constrictions. The challenge is to make such room-$T$ quantum dots with sufficient precision to obtain reproducible characteristics for different devices, which is hard to achieve by standard electron-beam lithography and isotropic dry etching.

An alternative route to graphene-based electronics is to consider graphene not as a new channel material for field-effect transistors (FET) but as a conductive sheet, in which various nm-size structures can be carved to make a single-electron-transistor (SET) circuitry. The idea is to exploit the fact that, unlike other materials, graphene nanostructures are stable down to true nm sizes and, possibly, even down to a single benzene ring. This allows one to explore a venue somewhere in between SET and molecular electronics (but by using the top-down approach). The advantage is that everything including conducting channels, quantum dots, barriers and interconnects can be cut out from a graphene sheet, whereas other material characteristics are much less important for the SET architecture[87,88] than for traditional FET circuits. This approach is illustrated in Fig. 6 that shows a SET made entirely from graphene by using electron-beam lithography and dry etching. For a minimum feature size of ≈10nm the combined Coulomb and confinement gap reaches ≈$10kT$, which should allow a SET-like circuitry operational at room $T$ (Fig. 6b), whereas resistive (rather than traditional tunnel) barriers can be used to induce Coulomb blockade. The SET architecture is relatively well developed[87,88], and one of the main reasons it failed to impress so far is difficulties with the extension of its operation to room $T$. The fundamental cause for the latter is a poor stability of materials for true-nm sizes, at which the Si-based technology is also to encounter fundamental limitations, according to the semiconductor industry roadmap. This is where graphene can come into play.

It is most certain that we will see many efforts to develop various approaches to graphene electronics. Whichever approach prevails, there are two immediate challenges. First, despite the recent progress in epitaxial growth of graphene,[33,34] high-quality wafers suitable for industrial applications still remain to be demonstrated. Second, one needs to control individual features in graphene devices accurately enough to provide sufficient reproducibility in their properties. The latter is exactly the same challenge that the Si technology has been dealing with successfully. For the time being, proof-of-principle nm-size graphene devices can be made by electrochemical etching using scanning-probe nanolithography.[89]

GRAPHENE DREAMS
Despite the reigning optimism about graphene-based electronics, "graphenium" microprocessors are unlikely to appear for the next 20 years. In the meantime, one can certainly hope for many other graphene-based applications to come of age. In this respect, clear parallels with nanotubes allow a highly educated guess of what to expect soon.

The most immediate application for graphene is probably its use in composite materials. Indeed, it has been demonstrated that a graphene powder of uncoagulated micron-size crystallites can be produced in a way scaleable to mass production[17]. This allows conductive plastics at less than 1 volume percent filling[17], which in combination with low production costs makes graphene-based composite materials attractive for a variety of uses. However, it seems doubtful that such composites can match the mechanical strength of their nanotube counterparts because of much stronger entanglement in the latter case.

Another enticing possibility is the use of graphene powder in electric batteries that are already one of the main markets for graphite. An ultimately large surface-to-volume ratio and high conductivity provided by graphene powder can lead to improvements in batteries' efficiency, taking over from carbon nanofibres used in modern batteries. Carbon nanotubes have also been considered for this application but graphene powder has an important advantage of being cheap to produce[17].

One of the most promising applications for nanotubes is field emitters and, although there have been no reports yet about such use of graphene, thin graphite flakes were used in plasma displays (commercial prototypes) long before graphene was isolated, and many patents were filed on this subject. It is likely that graphene powder can offer even more superior emitting properties.

Carbon nanotubes were reported to be an excellent material for solid-state gas sensors but graphene offers clear advantages in this particular direction[41]. Spin-valve and superconducting field-effect transistors are also obvious research targets, and recent reports describing a hysteretic magnetoresistance[90] and substantial bipolar supercurrents[91] prove graphene's major potential for these applications. An extremely weak spin-orbit coupling and the absence of hyperfine interaction in $^{12}C$-graphene make it an excellent if not ideal material for making spin qubits. This guarantees graphene-based quantum computation to become an active research area. Finally, we cannot omit mentioning hydrogen storage, which has been an active but controversial subject for nanotubes. It has already been suggested that graphene is capable of absorbing an ultimately large amount of hydrogen[92], and experimental efforts in this direction are duly expected.

AFTER THE GOLD RUSH

It has been just over 2 years since graphene was first reported and, despite remarkably rapid progress, only the very tip of the iceberg has been uncovered so far. Because of the short time scale, most experimental groups working now on graphene have not published even a single paper on the subject, which has been a truly frustrating experience for theorists. This is to say that, at this time, no review can possibly be complete. Nevertheless, the research directions explained or pencilled here should persuade even die-hard sceptics that graphene is not a fleeting fashion but is here to stay, bringing up both more exciting physics and, perhaps, even wide-ranging applications.

**Acknowledgements**. We are most grateful to Irina Grigorieva, Alberto Morpurgo, Uli Zeitler, Antonio Castro Neto, Allan MacDonald and Fabio Pulizzi for many useful comments that helped to improve this review.